\documentstyle[11pt]{article}
\newcommand{\noi}{\vspace{12pt}\noindent}

\newcommand{\beq}{\begin{equation}}
\newcommand{\eeq}{\end{equation}}
\newcommand{\bea}{\begin{eqnarray}}
\newcommand{\eea}{\end{eqnarray}}
\newcommand{\lpart}{\raise.3ex\hbox{$\stackrel{\leftarrow}{\partial}$}}
\newcommand{\rpart}{\raise.3ex\hbox{$\stackrel{\rightarrow}{\partial}$}}
\newcommand{\ldr}{\raise.3ex\hbox{$\stackrel{\leftarrow}{\delta}$}}
\newcommand{\deder}[1]{{ 
 {\stackrel{\raise.1ex\hbox{$\leftarrow$}}{\delta}   } 
\over {   \delta {#1}}  }}
\newcommand{\dedel}[1]{{ 
 {\stackrel{\lower.3ex \hbox{$\rightarrow$}}{\delta}   }
 \over {   \delta {#1}}  }}

\newcommand{\papar}[1]{{ 
 {\stackrel{\raise.1ex\hbox{$\leftarrow$}}{\partial}   } 
\over {   \partial {#1}}  }}
\newcommand{\papal}[1]{{ 
 {\stackrel{\lower.3ex \hbox{$\rightarrow$}}{\partial}   }
 \over {   \partial {#1}}  }}
\newcommand{\zzzz}{z}

\newcommand{\ad}{\mbox{\rm ad}}

\begin{document}
\thispagestyle{empty}
\vspace{3cm}
\title{\Large{\bf Hamiltonian N=2 Superfield Quantization}}
\vspace{2cm}
\author{{\sc I.A.~Batalin}\\
Lebedev Physics Institute\\
53 Leninisky Prospect\\Moscow 117924\\Russia\\
~\\and\\~\\
{\sc P.H.~Damgaard}\\The Niels Bohr Institute\\Blegdamsvej 17\\
DK-2100 Copenhagen\\Denmark}
\maketitle
\begin{abstract}
We present a superfield construction of Hamiltonian quantization
with $N=2$ supersymmetry generated by two fermionic charges
$Q^a$. As a byproduct of the analysis we also derive
a classically localized path integral from two
fermionic objects $\Sigma^a$ that can be viewed as ``square
roots'' of the classical bosonic action under the product of
a functional Poisson bracket.  
\end{abstract}

\vfill
\newpage



\noi
In two earlier papers \cite{super,supercurved}, it has been shown that 
BRST symmetry can be embedded in an $N=1$ superfield formalism of 
unconstrained superfields. This holds in both the  
Hamiltonian operator language and in the phase space path integral form,
and even in the general case when phase space is curved. 
Quite remarkably, also in the presence of second-class constraints
the appropriate superfield phase space path integral precisely
provides the correct Fradkin-Senjanovic path integral measure 
\cite{FradSen} after integrating
over the superfield partners of the ordinary fields \cite{supercurved}.
Related superfield formulations have later appeared in other
contexts as well \cite{other}.\footnote{Superfield formulations of the
Lagrangian antifield formalism \cite{BV} have also been considered
\cite{superBV,super}.} 

\noi
As expected, the required superspace is in that case
two-dimensional, spanned by ordinary time $t$ and a new
(real) fermionic direction denoted by $\theta$. All original 
phase space coordinates $z^A_0(t)$ are then extended to superfield phase space 
coordinates in the obvious manner
\beq
z^A(t,\theta) ~=~ z^A_0(t) + \theta z^A_1(t) ~,
\label{z}
\eeq
from which it follows that the superfield
$z^A(t,\theta)$ is of same Grassmann parity $\epsilon_A$ as $z^A_0(t)$.

\noi
In this paper we will show how to generalize a similar superfield
construction to the case of $N=2$ supersymmetry generated by two
fermionic charges $Q^a, a = 1,2$. 

\noi
$N=2$ superfield phase space variables have the following expansion,
\bea
z^A(t,\theta) &=& z_0^A(t) + \theta_az^{aA}(t) + 
\frac{1}{2}\epsilon^{ab}\theta_a\theta_bz^{3A}(t) \cr
&=& z_0^A(t) + \theta_az^{aA}(t) + \delta^{(2)}(\theta)z^{3A}(t) ~. 
\label{sp2z}
\eea
It follows that Grassmann parities are
\beq
\epsilon(z^A) = \epsilon(z^A_0) = \epsilon_A ~~,~~~
\epsilon(z^{aA}) = \epsilon_A+1 ~~,~~~ \epsilon(z^{3A}) = \epsilon_A ~.
\eeq
All other unconstrained $N=2$ superfields $F$ will have analogous
expansions that truncate at the top component $F^{3}$ as in eq.
({\ref{sp2z}). The zero-components $z_0^A(t)$ are identified with
the original phase space variables.

\noi
The graded Poisson bracket for superfields is defined by
\beq
\{F\left(\zzzz(t,\theta)\right),G\left(\zzzz(t,\theta)\right)\} 
~\equiv~ F\lpart_{A}\omega^{AB}\rpart_{B} G ~,\label{pb}
\eeq
with a symplectic superfield metric
\beq
\omega^{AB}\left(z(t,\theta)\right) ~=~ \{z^A(t,\theta),z^B(t,\theta)\} ~,
\eeq
that may or may not depend on $z(t,\theta)$. As is well known,
$\omega^{AB}$ has the following symmetry properties:
\beq
\omega^{AB} ~=~ - (-1)^{\epsilon_{A}\epsilon_{B}}\omega^{BA} ~~,~~~~~~
\epsilon(\omega^{AB}) = \epsilon_A + \epsilon_B ~,
\eeq
which implies
\beq
\{F,G\} ~=~ -(-1)^{\epsilon(F)\epsilon(G)}\{G,F\} ~.
\eeq
The condition
\beq
\omega^{AD}\partial_D\omega^{BC}(-1)^{\epsilon_{A}\epsilon_{C}}~ + 
~{\mbox{\rm cyclic}} ~=~ 0 ~
\eeq
guarantees the super Jacobi identity
\beq
\{\{F,G\},H\}(-1)^{\epsilon(F)\epsilon(H)}~ + 
~{\mbox{\rm cyclic}} ~=~ 0 ~.\label{Jacobi}
\eeq
The inverse symplectic metric is denoted by $\omega_{AB}$ so that
$\omega^{AB}\omega_{BC} = \delta^A_{~C}$.

\noi
We now introduce a constant symmetric
(bosonic) metric $g^{ab} = g^{ba}$ and two fermionic parameters
$\theta_a$. Next, we define two superfield derivatives 
\beq
D^a ~\equiv~ \frac{\partial}{\partial\theta_a} + g^{ab}\theta_b
\frac{\partial}{\partial t} ~=~ 
\partial^a + \theta^a\partial_t ~, \label{sp2deriv}
\eeq
where we raise indices
with the help of the metric, $i.e.$, $\theta^a = g^{ab}\theta_b$.
The superspace derivatives (\ref{sp2deriv})
are fermionic, and one immediately sees that
\beq
[D^a,D^b] ~\equiv~ D^{\{a}D^{b\}} ~=~ 2g^{ab}\partial_t ~.
\eeq
We are thus naturally led to choose the metric $g^{ab}$ constant and 
invertible with $g^{ab}g_{bc} = \delta^a_c$. 
There are, however, still two distinct classes of metric, 
depending on the sign of $\det(g)$. When
$\det(g)$ is positive (for convenience normalized to unity) the metric can
be continuously deformed to the identity, while for negative
$\det(g)$ (again conveniently normalized to minus unity), this is not
possible. When $\theta_a$ carries non-trivial ghost number,
ghost number conservation requires $\det(g)=-1$. An example
of an $N=2$ formalism with ghost number has been given in
ref. \cite{Ore} for the case of Lagrangian Yang-Mills theory
gauge fixed with BRST and anti-BRST symmetry. Further
extension to higher $N$ and non-trivial geometry 
can also be considered \cite{Hull}.

\noi
We will seek for a superfield action whose equations of motion are
\beq
D^a z^A(t,\theta) ~=~ \{Q^a(z(t,\theta);t,\theta),z^A(t,\theta)\} ~,
\label{sp2eom}
\eeq
where the $Q^a$ are fermionic.
Applying $D^b$ from the left to this equation, and symmetrizing
in $a$ and $b$ we get
\beq
\dot{z}^A ~=~ -\{H,z^A\} \label{sp2H}
\eeq
where the superfield Hamiltonian $H$ is defined by
\beq
{\cal D}^{\{a}Q^{b\}}
+\{Q^a,Q^b\} ~=~ -2g^{ab}H ~,\label{sp2Hdef}
\eeq
and ${\cal D}^a$ is the {\em explicit} differentiation
analog of $D^a$. We make no restriction
to the case of explicit $t$ or $\theta$ independence of $Q^a$
(and hence $H$).

\noi
Multiplying eq. (\ref{sp2H}) from the left by $\theta^a$, and
employing eq. (\ref{sp2eom}), we find
\beq
\partial^az^A ~=~ \{\Omega^a,z^A\} ~,\label{sp2partiala}
\eeq
where we have defined 
\beq
\Omega^a(z(t,\theta);t,\theta) ~\equiv~ Q^a(z(t,\theta);t,\theta) +
\theta^aH(z(t,\theta);t,\theta) ~.\label{sp2Omega}
\eeq
The superspace evolution in $\theta_a$ is thus dictated by the
combination $\Omega^a$, while the corresponding evolution in time
$t$ is dictated by the superfield Hamiltonian $H$ of eq. (\ref{sp2Hdef}).
At this stage $\Omega^a$ and $H$ appear on
similar footing, and both are derived from the same fundamental objects $Q^a$.

\noi
The superfield integrability conditions
\bea
\left(\partial^a\partial_t - \partial_t\partial^a\right)
z^A(t,\theta) &=& 0 \cr
\partial^{\{a}\partial^{b\}}z^A(t,\theta) &=& 0 ~,
\eea
are satisfied if
\bea
\partial_t\Omega^a + \{\Omega^a,H\} + \partial^aH &=& 0 \label{sp2consist1}\\
\partial^{\{a}\Omega^{b\}} + \{\Omega^a,\Omega^b\} &=& 0 ~,
\label{sp2consist2}
\eea
where $\partial_t$ and $\partial^a$ here stand for explicit $t$ and
$\theta_a$ derivatives.

\noi
Evaluating eq. (\ref{sp2H}) at $\theta_a=0$, we find
\beq
\dot{z}_0^A(t) ~=~ -\{H_0,z_0^A(t)\} ~,\label{sp2eomz0}
\eeq
where
\beq
H_0(z_0;t) ~=~ \left.-\frac{1}{4}g_{ab}\left(\partial^{\{a}Q^{b\}}
+ \{Q^a,Q^b\}\right)\right|_{\theta=0} ~.\label{sp2H0def}
\eeq
Consider now
\beq
X^A(t,\theta) ~\equiv~ \dot{z}^A(t,\theta) + \{H,z^A(t,\theta)\} ~,
\eeq
and perform a rescaling in $\theta_a$,
\beq
\theta_a ~\to~ \theta_a' ~\equiv~ \alpha\theta_a ~.\label{sp2scaling}
\eeq
Then,
\beq
\frac{dX^A}{d\alpha} ~=~ \frac{d}{dt}\frac{dz^A}{d\alpha}
+ \left\{\frac{\partial H}{\partial\alpha},z^A\right\}
-\left\{\theta_a Q^a,\{z^A,H\}\right\} ~,
\eeq
where we have used eq. (\ref{sp2partiala}) and $\theta_a\theta^a=0$,
which implies
\beq
\frac{dz^A}{d\alpha} ~=~ \{\theta_a Q^a,z^A\} ~.
\eeq
Differentiating w.r.t. time $t$ gives
\bea
\frac{d}{dt}\frac{dz^A}{d\alpha} &=& \left\{\theta_a\partial_t Q^a,
z^A\right\} + \left(X^B + \{z^B,H\}\right)\partial_B\{\theta_a Q^a,z^A\} \cr
&=& \left\{\theta_a\partial_t Q^a,z^A\right\} + 
X^B\partial_B\{\theta_a Q^a,z^A\}
- \left\{H,\{\theta_a Q^a,z^A\}\right\} ~.
\eea
Combining these results, we get, successively, 
\bea
\frac{dX^A}{d\alpha} &=& X^B\partial_B\{\theta_a Q^a,z^A\}
+ \left\{\frac{\partial H}{\partial\alpha},z^A\right\}
+ \{\theta_a\partial_t Q^a,z^A\} \cr
&&+ \left\{\theta_a Q^a,\{H,z^A\}\right\} - 
\left\{H,\{\theta_a Q^a,z^A\}\right\} \cr
&=& X^B\partial_B\{\theta_a Q^a,z^A\} + \left\{\frac{\partial H}
{\partial\alpha} + \theta_a\partial_t Q^a + \{\theta_a Q^a,H\},z^A\right\} ~.
\label{sp2X}
\eea
We now use the consistency condition (\ref{sp2consist1}). 
Performing the rescaling (\ref{sp2scaling}) this condition reads
\beq
\frac{\partial H}{\partial\alpha}
   + \{\theta_a Q^a,H\}  + \theta_a\partial_t Q^a ~=~ 0 ~,
\eeq
which, when inserted into eq. (\ref{sp2X}), gives
\beq
\frac{dX^A}{d\alpha} ~=~ X^B\partial_B\{\theta_a Q^a,z^A\} ~=~
X^B\omega_{BC}\left\{z^C,\{\theta_a Q^a,z^A\}\right\} ~.
\eeq
This ordinary homogenous differential equation which governs the
$\alpha$ (and hence $\theta_a$) evolution of $X^A$ shows that if
we choose $X^A|_{\alpha=0} = 0$, we have $X^A(\alpha) = 0$ for all
$\alpha$. Thus, if the superfield equations of motion
\beq
\dot{z}^A(t,\theta) ~=~ -\{H,z^A(t,\theta)\}
\label{sp2zdot}
\eeq
hold for $\theta_a = 0$ (where they explicitly coincide with the
classical equations of motion for the physical phase space
variables $z_0^A$ and Hamiltonian $H_0$ according to eq.
(\ref{sp2eomz0})), they hold for all $\theta_a$. 
The Hamiltonian dynamics of the zero-sector can be ``lifted'' to
the $N=2$ superspace. Not surprisingly,
it is precisely the integrability condition (\ref{sp2consist1}) which
guarantees this.

\noi
The next step is to find a suitable action from which the
equations of motion (\ref{sp2eom}) can be derived.
Immediate candidates are the following two {\em fermionic} functionals,
\beq
\Sigma^a = \int\!dtd^2\theta~ \left[z^A(t,\theta)\bar{\omega}_{AB}
D^az^B(t,\theta)(-1)^{\epsilon_B} + Q^a(z(t,\theta),t,\theta))\right] ~,
\label{Sigmaaction}
\eeq
where
\beq
\bar{\omega}_{AB} ~\equiv~ \left(z^C\partial_C + 2\right)^{-1}\omega_{AB}
~=~ \int_0^1\! \alpha d\alpha~\omega_{AB}(\alpha z) ~.
\label{omegabar}
\eeq
Indeed, one can readily verify that variations of $\Sigma^a$
precisely generate the equations of motion (\ref{sp2eom}).
However, the objects $\Sigma^a$ being Grassmann numbers, we cannot 
exponentiate them to let them take the r\^{o}le of actions.  So although 
they lead to the desired equations of motion (\ref{sp2eom}), we must seek
alternatives.


\noi
As a first attempt,
consider the following action $S$, derived with the help of a
functional Poisson bracket:
\bea
S &~\equiv~& \frac{1}{4}g_{ab}\left\{\Sigma^b,\Sigma^a\right\} \cr
&=& \frac{1}{4}g_{ab}\Sigma^b\int\! \deder{z^B(t',\theta')} dt'd^2\theta'
~\Omega^{BA}(t',\theta';t,\theta)~dtd^2\theta\dedel{z^A(t,\theta)}
\Sigma^a ~,
\eea
based on an ultralocal $\Omega^{AB}$: 
\beq
\Omega^{AB}(t',\theta';t,\theta) ~=~ \omega^{AB}(z(t,\theta))
\delta(t'-t)\delta^{(2)}(\theta'-\theta) ~.
\eeq
Inserting the definition (\ref{Sigmaaction}), we find
\bea
S &=& \frac{1}{4}\int\!dtd^2\theta~\left(-D^bz^C\omega_{CB}+Q^b\lpart_B
\right)g_{ba}\omega^{BA}\left(\omega_{AD}D^az^D(-1)^{\epsilon_D}
+\partial_A Q^a\right) \cr
&=& \int\!dtd^2\theta~\left(-\frac{1}{4}D^bz^Bg_{ba}\omega_{BA}
D^az^A(-1)^{\epsilon_A} - H\right) ~,\label{Gozzi0}
\eea
where $H$ is as given in eq. (\ref{sp2Hdef}). Note that it is
$\omega_{AB}$, and not $\bar{\omega}_{AB}$,
which enters in the kinetic term. We next derive the
equations of motion:
\bea
0 ~=~ \frac{\delta S}{\delta z^B(t,\theta)} &=&
\frac{1}{2}D^bg_{ba}\omega_{BA}D^az^A(-1)^{\epsilon_A+\epsilon_B}\cr
&& - \frac{1}{4}D^bz^Cg_{ba}\partial_B
\omega_{CA}D^az^A(-1)^{\epsilon_A+\epsilon_B(
\epsilon_C+1)}-\partial_BH \cr
&=& \omega_{BA}\dot{z}^A - \partial_BH ~.
\eea
These are the equations of motion of the $N=1$ case, although now
expressed in terms of $N=2$ superfields.
Integrating up these equations of motion, 
we find that the action is classically equivalent to
\beq
S ~=~ \int\!dtd^2\theta\left[z^B\bar{\omega}_{BA}\dot{z}^A - H\right] ~,
\label{Gozzi1}
\eeq
with $\bar{\omega}_{AB}$ defined as in eq. (\ref{omegabar}). At this stage
we also note that the action (\ref{Gozzi0}) has an equivalent 
first-order formulation in terms of an additional
superfield $\lambda_a^A(t,\theta)$ with $\epsilon(\lambda_a^A) =
\epsilon_A+1$:
\beq
S ~=~ \int\!dtd^2\theta~\left[\lambda_a^Ag^{ab}\omega_{AB}
\lambda_b^B(-1)^{\epsilon_B} +
\lambda_a^A\left(\omega_{AB}D^az^B(-1)^{\epsilon_B} +
\partial_AQ^a\right)\right] ~.
\eeq
The fact that only the $N=1$ equations of motion appear from the
action (\ref{Gozzi0}) should not be a surprise. For a $2n$-dimensional
superfield phase space there is an obvious impossibility of deriving the $4n$
equations of motion (\ref{sp2eom}). In fact, this is the origin of
a serious problem with the action (\ref{Gozzi0}). To simplify the
discussion, let us consider the case where $\omega_{AB}$ is constant.
Expanding the superfield according to eq. (\ref{sp2z}), and performing
the $\theta_a$-integrations, we are left with
\bea
S &=& \int\! dt~\left[z^{3A}\left(\omega_{AB}\dot{z}_0^B
-\partial_A H_0(z_0)\right)\right. \cr
&&+ \left.\frac{1}{2}\epsilon_{ab}z^{aA}\omega_{AB}\dot{z}^{bB}
(-1)^{\epsilon_B}- \frac{1}{2}\epsilon_{ab}z^{aA}\rpart_A H_0 \lpart_B
z^{bB}(-1)^{\epsilon_B}\right] ~,
\eea
where $H_0$ is as defined in eq. (\ref{sp2H0def}).
The top component $z^{3A}$ of the phase space superfield $z^A$
has ended up playing the r\^{o}le of a Lagrange multiplier that
imposes the classical equations of motion for the original phase
space variables $z_0$ {\em as a $\delta$-function constraint in
the path integral}. The path integral has been
localized on just the classical trajectories. Indeed, the remaining
$z^{aA}$-integrations precisely conspire to provide for the
Jacobian that renders the partition function equal to unity. So
although we have achieved an $N=2$ superfield phase space
path integral formulation with correct equations of motion, the
price we have paid is total absence of quantum fluctuations.

\noi
Interestingly, a path integral based on the action (\ref{Gozzi0}) or 
equivalently (\ref{Gozzi1}), which trivializes
the path integral dynamics to the classical trajectories has earlier
been arrived at from an entirely different context\footnote{The
idea of an operator formulation of classical mechanics was apparently 
suggested by Koopman and von Neumann in 1931-32, see ref. \cite{Gozzi}.} 
by Gozzi et al.
\cite{Gozzi}. Here we see that this ``path integral for 
classical physics'' can be deduced from an underlying principle
of two superfield equations of motion, and two fermionic actions
$\Sigma^a$ that are effectively square roots (w.r.t. the
product induced by the ordinary functional superfield Poisson bracket) 
of the bosonic
action $S$.


\noi
We now present a classical (bosonic) action that leads to the
correct equations of motion, and which does not localize on the
classical trajectories. We first introduce some notation. Let
\beq
\tilde{\theta}_a ~\equiv~ g_{ab}\epsilon^{bc}\theta_c ~.
\eeq
It follows that
\beq
\frac{1}{2}\theta^a\tilde{\theta}_a ~=~ \delta^{(2)}(\theta)~~~~,
~~~~~~~{\mbox{\rm and}} ~~~~ \tilde{\theta}^a ~=~
\partial^a\delta^{(2)}(\theta) ~,
\eeq
while $\theta^a\theta_a = \tilde{\theta}^a\tilde{\theta}_a = 0$. 
Similarly, let us introduce two derivatives
\beq
D ~\equiv~ \theta_aD^a ~=~ \theta_a\partial^a ~~,~~~~~ \tilde{D} ~\equiv~ 
\tilde{\theta}_aD^a ~.
\eeq
We also define a covariant derivative
\beq
\nabla^a ~\equiv~ D^a - \ad ~Q^a ~,
\eeq
by means of which the proposed equations of motion (\ref{sp2eom})
take the compact form
\beq
\nabla^a z^A(t,\theta) ~=~ 0 ~. \label{sp2eom'}
\eeq 
Here we have introduced the adjoint action w.r.t. the super
Poisson bracket, $\ad ~F \equiv \{F,~\cdot~ \}$.
Similarly, we also define
\beq
\nabla ~\equiv~ D - \ad ~Q ~~~ , ~~~~~~ 
\tilde{\nabla} ~\equiv~  \tilde{D} - \ad ~\tilde{Q} ~,\label{nabladef}
\eeq
where $Q \equiv \theta_aQ^a$ and $\tilde{Q} \equiv \tilde{\theta}_aQ^a$.
An important property
of $\tilde{D}$ is antisymmetry under transposition
(conjugation), $\tilde{D}^T = -\tilde{D}$, while, as can be seen, 
$D$ is neither symmetric nor antisymmetric. This means that 
$\tilde{D}$ (and not $D$) is a natural derivative to introduce in the kinetic
term of an $N=2$ action. 

\noi
We propose the following action:
\bea
S &=& -\frac{1}{2}\int\! dtd^2\theta~ \left(z^A\bar{\omega}_{AB}
\tilde{D}z^B + \tilde{Q}\right) \cr
&&+ \int\! dtd^2\theta~ \left(\nabla z^A\right)h^{~B}_{A}\lambda_B
+ \int\! dt~ \Pi^A\int\! d^2\theta~ \lambda_A ~, \label{sp2S}
\eea
where in addition to the $2n$ phase space superfield variables
$z^A$ we have added $2n$ Lagrange superfield multipliers 
$\lambda_A(t,\theta)$ and Lagrange multipliers
$\Pi^A(t)$. We have also introduced the superfield vielbeins $h_A^{~B}$
of the symplectic metric, defined by
\beq
\omega_{AB} ~=~ (-1)^{\epsilon_B(1+\epsilon_D)}h_A^{~C}
\omega^{0}_{CD}h_B^{~D} ~,
\eeq
where $\omega^{0}_{AB}$ is the superfield symplectic metric in
Darboux form. With the vielbeins inserted in front of the Lagrange 
multiplier $\lambda_B$ in the second line of (\ref{sp2S}) 
that term shares the reparametrization invariance of the
rest of the action. We note that the vielbeins are invertible.

\noi
The variational equations of motion from (\ref{sp2S}) are
\beq
\lambda_A ~=~ 0 ~~,~~~~ \Pi^A ~=~ 0 ~,
\eeq
and hence, for the superfield phase space variables,
\beq
\tilde{\nabla} z^A ~=~ 0 ~, \label{nablatildez}
\eeq
and
\beq
\nabla z^A ~=~ 0 ~. \label{nablaz}
\eeq
As we shall show below, the two sets of equations (\ref{nablatildez})
and (\ref{nablaz}) are, taken together, equivalent
to the $N=2$ equations of motion (\ref{sp2eom}).

\noi
It is instructive to first view the action (\ref{sp2S}) in component form.
To simplify this component analysis let us note that if we are only 
interested in the classical
equations of motion, we can ignore the presence of vielbeins by formally
setting $h_A^{~B}  = \delta_A^{~B}$ in eq. (\ref{sp2S}). This is
because the vielbeins, being invertible, are only responsible for the
correct path integral measure which comes just from the integration over
$\lambda_A$. We also remind the reader that this Lagrange multiplier 
$\lambda_A$ has an expansion
\beq
\lambda_A(t,\theta) ~=~ \lambda_A^0 + \theta_a\lambda_A^a +
\frac{1}{2}\epsilon^{ab}\theta_a\theta_b\lambda_A^3 ~,
\eeq
but the last term in (\ref{sp2S}) simply removes the top component
$\lambda_A^3$\footnote{An alternative to introducing this term 
explicitly is to work with a constrained superfield $\lambda_A$ whose
top component is required to vanish identically. We prefer to use
the formulation with an unconstrained superfield.}. 
Expanding the rest by means of (\ref{sp2z}), and performing the 
$\theta^a$-integrations, we get:
\bea
S &=& \int\!dt~ \left[z_0^A\bar{\omega}_{AB}\dot{z}_0^B
+ \frac{1}{2}g_{ab}\partial^aQ^b + \frac{1}{4}z^{aA}\omega_{AB}g_{ab}
z^{Bb}(-1)^{\epsilon_B}\right. \cr
&&+ \frac{1}{2}g_{ab}z^{aA}\partial_AQ^b + \epsilon_{ab}
\left(z^{aA} - \{Q^a,z_0^A\}\right)\lambda_A^b(-1)^{\epsilon_A} \cr
&&+ 2\left.\left(z^{3A} - \frac{1}{2}\epsilon_{ab}\left(\{\partial^aQ^b,z_0^A\}
+ z^{aB}\partial_B\{Q^b,z_0^A\}\right)\right)\lambda_A^0\right] ~.
\label{sp2Scomp}
\eea
Varying the action (\ref{sp2Scomp}) w.r.t. $z^{3A}$ we indeed verify
that on-shell $\lambda_A^0 = 0$, while varying w.r.t. $\lambda_A^a$
then yields
\beq
z^{aA} ~=~ \{Q^a,z_0^A\} ~. \label{zaAeq}
\eeq
Thus, the fields $z^{aA}$ are not independent, but given by
the $Q^a$-transform of the original phase space variables $z_0$.
In turn, varying with respect to $z^{aA}$ and using eq. (\ref{zaAeq}) 
we verify that $\lambda^{a}_A = 0$.
Similarly for $z^{3A}$: varying w.r.t. $\lambda_A^0$ and using 
(\ref{zaAeq}), we get
\beq
z^{3A} ~=~ \frac{1}{2}\epsilon_{ab}\left(\{\partial^aQ^b,z_0^A\} +
\{Q^a,\{Q^b,z_0^A\}\}\right) ~.
\eeq
Inserting the above identifications back into (\ref{sp2Scomp}), 
we find that at the
classical level this action is equivalent to
\beq
S = \int\!dt~ \left[z_0^A\bar{\omega}_{AB}\dot{z}_0^B
+ \frac{1}{2}g_{ab}\partial^aQ^b -\frac{1}{4}
\{Q^a,z_0^A\}\omega_{AB}g_{ab}\{z_0^B,Q^b\}
+ \frac{1}{2}g_{ab}\{Q^a,z_0^A\}\partial_AQ^b\right] ~.
\eeq
The last three terms neatly conspire to yield
\beq
S = \int\!dt~ \left(z_0^A\bar{\omega}_{AB}\dot{z}_0^B
-H_0\right) ~,
\eeq
where $H_0$ is as defined in eq. (\ref{sp2H0def}). 
This is just the phase space action needed for the original
phase space variables $z_0$ associated with the classical
Hamiltonian $H_0$, and equations of motion
\beq
\dot{z}_0^A ~=~ -\{H_0,z_0^A\} ~.
\label{sp2z0dot}
\eeq
Remarkably, due to the explicit presence of
the vielbein $h_A^{~B}$ in eq. (\ref{sp2S}), integrations over
$\Pi^A,\lambda_A,z^{aA}$ and $z^{3A}$ all precisely combine to
yield the required measure factor Pf$(\omega(z_0))$ in the BFV path
integral over the remaining phase space variables $z_0$. This is 
precisely as in the $N=1$ case \cite{super}.

\noi
Our next aim is to show that the two equations (\ref{nablatildez})
and (\ref{nablaz}) in fact are equivalent to the equations of
motion (\ref{sp2eom'}). Of course, the opposite statement is trivially
true: we recover (\ref{nablatildez}) and (\ref{nablaz}) by 
multiplying eq. (\ref{sp2eom'}) by $\tilde{\theta}_a$ and 
$\theta_a$, respectively. Let us now, conversely, consider
eq. (\ref{nablatildez}). Using the definition (\ref{nabladef})
we conclude that
\bea
\nabla^az^A &=& g^{ab}\epsilon^{cd}\theta_dF^A_{bc} \cr
&=&\frac{1}{2}\epsilon^{ad}\theta_dg^{bc}F^A_{bc}
+ \frac{1}{2}\theta_dg^{\{ab}\epsilon^{cd\}}F^A_{bc}\label{Feq}
\eea
where $F^A_{ab}$ is a so far undetermined (superfield) function
which is symmetric in the lower indices,
$F^A_{ab} = F^A_{ba}$, and where in the second line we have
split up in symmetric and antisymmetric parts in indices $a$ and $d$.
Multiplying eq. (\ref{Feq}) by $\theta_a$ from the left, and
making use of eq. (\ref{nablaz}), gives
\beq
\delta^{(2)}(\theta)g^{bc}F^A_{bc} ~=~ 0 ~,
\eeq
from which we conclude that
\beq
g^{bc}F^A_{bc} ~=~ - \theta_e E^{eA} ~,\label{FEeq}
\eeq
for a new superfield function $E^{eA}$. Next, apply $D^a$ on
eq. (\ref{Feq}) and use (\ref{sp2deriv}) as well as the 
equations of motion ``lifted'' to superspace, eq. (\ref{sp2zdot}),
to get
\beq
g^{\{ac}\epsilon^{db\}}F^A_{cd} ~=~ -\theta_eG^{eabA} ~,\label{FGeq}
\eeq
where $G^{eabA} = G^{ebaA}$ is given by
\beq
G^{eabA} ~\equiv~ g^{\{bc}\epsilon^{de}F^B_{cd}\partial_B
\{Q^{a\}},z^A\} - g^{\{ac}\epsilon^{de}D^{b\}}F^A_{cd} ~.
\eeq
Logically, we have the right to
use eq. (\ref{sp2zdot}) at this stage as eqs.
(\ref{nablatildez}) and (\ref{nablaz}) themselves imply the 
zero-sector dynamics (\ref{sp2eomz0}) and, hence, 
as we have shown above in eqs.(\ref{sp2eomz0})-(\ref{sp2zdot}), 
the "lifted" dynamics (\ref{sp2zdot}). 
Substituting (\ref{FEeq}) and 
(\ref{FGeq}) into (\ref{Feq}), we conclude that
\beq
\nabla^a z^A ~=~ \delta^{(2)}(\theta)I^{aA} ~, \label{Ieq}
\eeq
for yet another superfield function $I^{aA}$. This in
turn, by the same argument that led to eq. (\ref{FGeq}), implies
\beq
\epsilon^{\{bc}\theta_cI^{a\}A} ~=~ -\delta^{(2)}(\theta)K^{baA}
\eeq
for a superfield function $K^{abA} = K^{baA}$.  Thus, 
$I^{aA}$ can have no zero-component, which, when plugged into
eq. (\ref{Ieq}) finally gives
\beq
\nabla^az^A ~=~ 0 ~,
\eeq
as we wished to show. The impossibility of having an action
that depends on only $2n$ phase space variables giving rise to
the $4n$ $N=2$ equations of motion (\ref{sp2eom'}) is circumvented
by splitting up these $4n$ equations into $2n$ equations of
motion (through eq. (\ref{nablatildez})) and $2n$ constraints
(through (\ref{nablaz})) whose r\^{o}le in addition is to assure
that the superfield partners of the original phase space
variables are given by canonical transformations. 

\noi
We finally remark that also at the operator level the two equations
\beq
\tilde{\nabla} \hat{z}^A(t,\theta) ~=~ 0 ~~,~~~~
\nabla \hat{z}^A(t,\theta) ~=~ 0 \label{nablatildeQM}
\eeq
can be shown to be equivalent to the $N=2$
superfield quantum equations of motion
\beq
\nabla^a \hat{z}^A(t,\theta) ~=~ 0 ~.\label{nablaQMeom}
\eeq
Here
\beq
\nabla^a ~\equiv~ D^a - (i\hbar)^{-1}\ad~\hat{Q}^a \label{nablaQM}
\eeq
with $\ad~\hat{F} \equiv [\hat{F},~\cdot~]$ denoting the adjoint action 
of the operator $\hat{F}$, and with the operators $\nabla$ and 
$\tilde{\nabla}$ being defined as in eq. (\ref{nabladef}) with
the obvious replacements of Poisson brackets with commutators.
Moreover, the two equations (\ref{nablatildeQM}), or, 
equivalently, the equations of motion (\ref{nablaQMeom})
assure compatibility between the fundamental equal-time commutation relation
\beq
[\hat{z}_0^A(t),\hat{z}_0^B(t)] ~=~ i\hbar\hat{\omega}(\hat{z}_0(t))
\eeq
and this commutation relation lifted to the equal-$t$ and equal-$\theta_a$
superspace commutation relation,
\beq
[\hat{z}^A(t,\theta),\hat{z}^B(t,\theta)] ~=~ 
i\hbar\hat{\omega}(\hat{z}(t,\theta)) ~.
\eeq
As a special case, we can consider the Hamiltonian
superfield for theories with 1st class constraints
and $Sp(2)$ symmetry,
a long sought-for generalization of the superfield formalism
for BRST symmetry \cite{super}
in which the two charges $\hat{Q}^a$ appear on equal footing. 
That case corresponds, in the operator formalism,
to the algebra-generating condition
$[\hat{Q}^a,\hat{Q}^b] = 0$. Introducing a ``gauge-fixing boson'' $F$
we can explicitly construct an $Sp(2)$ invariant superfield 
unitarizing Hamiltonian by means of the substitution
\beq
\hat{Q}^a ~\to~ exp\left[(i\hbar)^{-2}\theta_bg^{bc}\epsilon_{cd}~\ad
\left([\hat{Q}^d,\hat{F}]\right)\right]\hat{Q}^a ~.
\eeq
This, and other aspects of the present
superfield formalism, will be discussed elsewhere \cite{BD}.

\noi
To conclude, we have shown how to formulate $N=2$ superfield
Hamiltonian dynamics on a three-dimensional superspace 
spanned by time $t$ and two fermionic directions $\theta_a$.
The starting point of our construction
is a set of two fermionic charges $Q^a$ and two superspace
derivatives $D^a$. From a combination of the $Q^a$'s we derive
an Hamiltonian which governs the time
evolution of the superfield phase space variables, and whose
$\theta_a=0$ part gives the Hamiltonian of the
original phase space variables. From $Q^a$ also follows two fermionic
charges $\Omega^a$ which generate translations in the two
$\theta_a$-directions. A superfield phase space path integral 
for this $N=2$ theory has been proposed, and shown to reduce to
the usual phase space path integral upon integration 
over the $\theta_a$-variables and after integrating out all
auxiliary variables of the path integral. Remarkably, even
the correct path integral measure with the Pf($\omega(z_0)$)-factor
comes out automatically, thus generalizing the result of ref.
\cite{super} to this setting.

\vspace{1cm}
\noindent
{\sc Acknowledgement:}~ This work has been partially supported by
INTAS grant 00-0262. Also, the work of I.A.B has been partially supported
by President grant 00-15-96566 and RFBR grant 02-01-00930.
I.A.B. would like to thank the Niels Bohr Institute 
for the warm hospitality extended to him there, and both authors thank 
Klaus Bering for numerous discussions at an early stage of this work.

 \end{document}